\documentclass[aps,prd,amssymb,twocolumn,superscriptaddress,floatfix,nofootinbib,10pt]{revtex4-2}
\usepackage{setspace}
\usepackage[utf8]{inputenc}
\usepackage{float}
\usepackage{amsmath,amssymb,amsfonts,bm}
\usepackage{graphicx}
\usepackage{multirow}
\usepackage[usenames,dvipsnames]{xcolor}
\allowdisplaybreaks
\usepackage[colorlinks,breaklinks,linkcolor=Blue,citecolor=Green,urlcolor=NavyBlue]{hyperref}

\usepackage{appendix}
\usepackage{orcidlink}
\usepackage{subfigure}
\usepackage{soul}
\usepackage{physics}
\usepackage{booktabs}
\usepackage{ulem}
\usepackage{tikz-feynman}

\usepackage[cochineal,bigdelims,vvarbb]{newtxmath}
\usepackage[T1]{fontenc}

\newcommand{\ols}[1]{\mskip.5\thinmuskip\overline{\mskip-.5\thinmuskip {#1} \mskip-.5\thinmuskip}\mskip.5\thinmuskip} 
\newcommand{\olsi}[1]{\,\overline{\!{#1}}} 

\usepackage[compact]{titlesec}
\titleformat{\section}
   {\centering\normalfont\bfseries\MakeUppercase}{\thesection}{1em}{}
\titlespacing*{\section}
{0pt}{3ex plus 1ex minus 1ex}{0ex plus 1ex}
\titlespacing*{\subsection}
{0pt}{2ex plus 1ex minus 1ex}{0.5ex plus 0.5ex}

\allowdisplaybreaks

\begin{document}





\title{\boldmath Signatures of the $\Omega(2012)^{-}$ state in $\Xi^*\olsi{K}$ Correlation Functions}

\author{Jia-Xin Lin\,\orcidlink{https://orcid.org/0009-0006-6382-1035}}
\email[]{linjx@seu.edu.cn}
\affiliation{School of Physics, Southeast University, Nanjing 210094, China}%
\affiliation{Instituto de F\'{i}sica Corpuscular, Centro Mixto Universidad de Valencia-CSIC, Institutos de Investigaci\'{o}n de Paterna, Aptdo. 22085, E-46071 Valencia, Spain}

\author{P. Encarnación\orcidlink{0009-0005-0749-3885}}
	\email{Pablo.Encarnacion@ific.uv.es}
	\affiliation{Departamento de F\'{i}sica Teórica and IFIC, Centro Mixto Universidad de Valencia-CSIC, Institutos de Investigaci\'{o}n de Paterna, Aptdo. 22085, E-46071 Valencia, Spain}

\author{A. Feijoo\orcidlink{0000-0002-8580-802X}}
        \email{edfeijoo@ific.uv.es}
	\affiliation{Instituto de F\'{i}sica Corpuscular, Centro Mixto Universidad de Valencia-CSIC, Institutos de Investigaci\'{o}n de Paterna, Aptdo. 22085, E-46071 Valencia, Spain}
    
\author{M. Albaladejo\orcidlink{0000-0001-7340-9235}}
	\email{Miguel.Albaladejo@ific.uv.es}
	\affiliation{Instituto de F\'{i}sica Corpuscular, Centro Mixto Universidad de Valencia-CSIC, Institutos de Investigaci\'{o}n de Paterna, Aptdo. 22085, E-46071 Valencia, Spain}
 
\date{\today}

\begin{abstract}
We investigate the $\Omega(2012)$ resonance in the strangeness $S=-3$ sector within a coupled-channel chiral unitary approach and present the first quantitative predictions for femtoscopic correlation functions directly sensitive to its dynamics. The $\Omega(2012)$ is dynamically generated as a quasi-bound $\Xi^{\ast}\olsi{K}$--$\Omega\eta$ molecular state, with its coupling to the $\Xi\bar{K}$ channel driven by $d$-wave transitions. Model parameters are constrained by the measured mass, width, and the Belle determination of the branching fraction $\mathcal R^{\Xi\olsi{K}\pi}_{\Xi\olsi{K}}$, yielding $M_{\Omega(2012)}=(2012.53\pm0.73)$~MeV and $\Gamma_{\Omega(2012)}=(4.05\pm0.13)$~MeV.

Within this framework, we compute the femtoscopic correlation functions of the $\Xi^{\ast0}K^-$, $\Xi^{\ast-}\olsi{K}{}^0$, and $\Omega^-\eta$ systems. The $\Xi^{\ast}\olsi{K}$ correlation functions exhibit pronounced near-threshold structures that arise from the proximity of the $\Omega(2012)$ pole, demonstrating an exceptional sensitivity to its position and coupled-channel composition. In particular, the $\Xi^{\ast0}K^-$ correlation function is identified as a clean and highly selective probe of the $\Omega(2012)$ resonance. These results establish femtoscopic correlation measurements as powerful tools for extracting resonance properties beyond conventional invariant-mass analyses and provide concrete theoretical benchmarks for upcoming experimental studies aimed at elucidating the molecular nature of the $\Omega(2012)$.

\end{abstract}

\maketitle

\hyphenation{BESIII} 

\section{Introduction}\label{sec:Intr}

A recent high-statistics analysis by the ALICE Collaboration revealed a 15$\sigma$-significance signal near $2013$~MeV in the $K^{0}_S\Xi^-$ invariant-mass distribution from $pp$ collisions at $\sqrt{s}=13$~TeV~\cite{ALICE:2025atb}, marking the third independent confirmation of the $\Omega(2012)^-$ resonance. This observation was shortly preceded by evidence reported for a production mechanism of the $\Omega(2012)^-$ via the process $e^+e^-\to \Omega(2012)^- \ols{\Omega}{}^{+} + \text{c.c.}$, with a weaker significance of 3.5$\sigma$~\cite{BESIII:2024eqk}. Interestingly, the BESIII Collaboration also provided evidence for a new state, the $\Omega(2109)^-$ hyperon, produced through the same mechanism. Almost simultaneously, Belle presented improved measurements of the $\Omega(2012)^-$ properties and the branching fraction $\mathcal{R}^{\Xi\olsi{K}\pi}_{\Xi\olsi{K}}$~\cite{Belle:2022mrg}, incorporating refined $\Xi(1530) [\equiv \Xi^{\ast}]$ modeling and three-body phase-space effects, yielding a branching fraction value of $\mathcal{R}^{\Xi\olsi{K}\pi}_{\Xi\olsi{K}}=0.99 \pm 0.26 \pm 0.06$. 

The Belle analyses \cite{Belle:2019zco,Belle:2022mrg} were the natural follow-up stages to the original discovery of the $\Omega(2012)^-$ resonance \cite{Belle:2018mqs}, observed in its decay channels $K^{0}_S\Xi^-$, $K^{-}\Xi^0$, and their charge conjugates, using $e^+e^-$ collision data collected near the $\Upsilon(1S)$, $\Upsilon(2S)$ and $\Upsilon(3S)$ resonances. In that initial experimental analysis \cite{Belle:2018mqs}, the $\Omega(2012)^-$ spin-parity was inferred to be most likely $J^P=3/2^-$, based on comparisons with earlier Lattice QCD (LQCD) results \cite{Engel:2013ig} and various predictions from the Constituent Quark Model (CQM) approach \cite{Capstick:1986ter,Loring:2001ky,Oh:2007cr,Pervin:2007wa,Faustov:2015eba}. This assignment is further supported by the measured decay width, approximately 6 MeV, which is consistent with a $d$-wave decay of a $J=3/2$ state, whereas an $s$-wave decay of a $J=1/2$ state would be expected to result in a significantly broader width. 

To underscore the importance of these experimental breakthroughs, it is worth noting that, for nearly four decades, only three excited  $\Omega^*$ states---$\Omega(2250)$, $\Omega(2380)$ and  $\Omega(2470)$---with two- or three-star status were listed in the Review of Particle Physics (RPP) by the Particle Data Group (PDG) \cite{ParticleDataGroup:2018ovx}. These recent experimental efforts have therefore significantly refined our understanding of excited $\Omega^*$ baryon spectroscopy---which still remains sparse compared to that of other light baryons---and provided critical input for theoretical models.

In order to determine the elusive excited $\Omega^*$ states missing in this baryon spectrum, several such states had been predicted within various theoretical approaches---ranging from quark \cite{Capstick:1986ter,Loring:2001ky,Pervin:2007wa,Faustov:2015eba,Chao:1980em,Kalman:1982ut,An:2013zoa,An:2014lga} and Skyrme models \cite{Oh:2007cr} to LQCD simulations \cite{Engel:2013ig,CLQCD:2015bgi}---prior to the first evidence of the $\Omega(2012)$ \cite{Belle:2018mqs}. Following the initial measurement by the Belle Collaboration and its subsequent confirmations, a plethora of theoretical studies have addressed the nature of the $\Omega(2012)^{-}$, mainly supporting two possible interpretations: either as an excited conventional baryon or as a dynamically generated molecular state. The works based on CQM approaches~\cite{Xiao:2018pwe,Aliev:2018syi,Aliev:2018yjo,Wang:2018hmi,Polyakov:2018mow,Liu:2019wdr,Liu:2020yen,Arifi:2022ntc,Zhong:2022cjx,Wang:2022zja,Luo:2025cqs} interpret this state most likely as $P$-wave excitation, while other groups~\cite{Valderrama:2018bmv,Lin:2018nqd,Pavao:2018xub,Huang:2018wth,Gutsche:2019eoh,Lu:2020ste,Ikeno:2020vqv,Lin:2019tex,Ikeno:2022jpe,Lu:2022puv,Han:2025gkp,Shen:2025xcq}, motivated by the proximity of the nominal $\Omega(2012)^{-}$ mass to the $\olsi{K}\Xi^{\ast}$ threshold, favor a dynamically generated molecular interpretation within a coupled-channel framework. The latter description was already proposed in Refs.~\cite{Sarkar:2004jh,Hofmann:2006qx} before the first $\Omega(2012)^{-}$ observation \cite{Belle:2018mqs}. After the controversy sparked by the misleading estimation of $\mathcal{R}^{\Xi\olsi{K}\pi}_{\Xi\olsi{K}}<11.9\%$~\cite{Belle:2019zco} was resolved in Ref.~\cite{Belle:2022mrg}, the updated branching fraction renders the molecular interpretation the favored one. This outcome has further motivated a recent study~\cite{Li:2026wck}, which postulates the $\Omega(2380)$ resonance as a partner state of the $\Omega(2012)$, originating from the $\bar{K}^*\Xi^*$ -- $\omega\Omega$ -- $\phi\Omega$ coupled-channel dynamics.

With the recent reconstruction of the $\Omega(2012)^{-}$ resonance~\cite{ALICE:2025atb} and the measurement of $K^+\Xi^-$ correlation function (CF)~\cite{ALICE:2025kma} by the ALICE Collaboration, the experimental conditions are now in place for a potential measurement of the $\bar{K}\Xi(1530)$ CF---a system intimately connected to the $\Omega(2012)^{-}$ structure. A model-independent inverse-method analysis of such femtoscopic data, as performed in Refs.\,\cite{Ikeno:2023ojl,Feijoo:2023sfe,Albaladejo:2023wmv}, could shed further light on the presumably dominant molecular nature of the $\Omega(2012)^{-}$ through the extraction of its compositeness with respect to the relevant coupled channels, as well as the corresponding scattering parameters. 
These quantities could then be compared with theoretical predictions from different models to provide a more decisive discrimination among them. As a natural first step, a theoretical prediction of the $\bar{K}\Xi(1530)$ CF could serve as guidance for experimental analyses. Moreover, a future comparison to such a measurement would offer crucial information to refine the dynamical ingredients implemented in the different theoretical models, leading to an improved description of the $\Omega(2012)$ state. In this context, femtoscopy has been identified as a key tool to probe the nature of exotic hadrons, as reviewed in Ref.~\cite{Liu:2024uxn}, where it is shown how information can be extracted both from the shape of correlation functions and from the interaction encoded in the scattering wave function.

The present study provides the first quantitative prediction of the $\bar{K}\Xi(1530)$ correlation function based on a realistic interaction model consistent with the $\Omega(2012)$ phenomenology. This offers a direct theoretical benchmark for forthcoming femtoscopic measurements at the LHC.  

\section{Formalism}\label{sec:Form}
We employ the Koonin–Pratt (KP) formalism~\cite{Koonin:1977fh,Pratt:1990zq,Bauer:1992ffu} to compute the $\bar{K}\Xi(1530)$ CF, incorporating all relevant dynamical ingredients for a consistent theoretical description. The scattering wave function is obtained within a coupled-channel chiral unitary approach, based primarily on the framework of Refs.~\cite{Pavao:2018xub,Ikeno:2020vqv}, with the modifications detailed below.

\subsection{\boldmath $\Omega(2012)^-$ as a dynamically generated state}

In this framework, the $\Omega(2012)^-$ emerges as a $\Xi^\ast\olsi{K}$-$\Omega\eta$ quasi-bound state generated from meson–baryon dynamics. Following Refs.~\cite{Pavao:2018xub,Ikeno:2020vqv}, we consider the coupled channels $\Xi^\ast\olsi{K}$, $\Omega\eta$, and $\Xi\olsi{K}$, using the corresponding physical charge states $\Xi^{\ast0}K^-$, $\Xi^{\ast-}\olsi{K}{}^0$, $\Omega^-\eta$, $\Xi^0K^-$, and $\Xi^-\olsi{K}{}^0$, with thresholds ranging from $1808$ to $2220\,\text{MeV}$. 
%
%

Then, the transition potential in the physical basis using the conventions $(\olsi{K}{}^0, -K^-)$, $(\Xi^0, -\Xi^-)$, and $(\Xi^{\ast0}, \Xi^{\ast-})$, calculated from the isospin components as in Ref.~\cite{Sarkar:2004jh,Pavao:2018xub}, takes the form
\begin{equation}\label{eq:Vmatrix}
\setlength{\arraycolsep}{2pt}
    V =
\left(\begin{array}{c|ccccc}
   & \Xi^{\ast 0}K^- & \Xi^{\ast -}\olsi{K}{}^0 & \Omega^-\eta & \Xi^0K^- & \Xi^-\olsi{K}{}^0
   \\
   \hline
   \Xi^{\ast 0}K^- &
   -F & F & -\tfrac{3}{\sqrt{2}}F & \tfrac{1}{2}\alpha q_{l}^{2} & -\tfrac{1}{2}\alpha q_{l}^{2}
   \\
   \Xi^{\ast -}\olsi{K}{}^0 &
   F & -F & -\tfrac{3}{\sqrt{2}}F & \tfrac{1}{2}\alpha q_{l}^{2} & -\tfrac{1}{2}\alpha q_{l}^{2}
   \\
   \Omega^-\eta &
   -\tfrac{3}{\sqrt{2}}F & -\tfrac{3}{\sqrt{2}}F & 0 & -\tfrac{1}{\sqrt{2}}\beta q_{l}^{2} & \tfrac{1}{\sqrt{2}}\beta q_{l}^{2}
   \\
   \Xi^0K^- &
   \tfrac{1}{2}\alpha q_{l}^{2} &
   \tfrac{1}{2}\alpha q_{l}^{2} &
   -\tfrac{1}{\sqrt{2}}\beta q_{l}^{2} &
   0 & 0
   \\
   \Xi^-\olsi{K}{}^0 &
   -\tfrac{1}{2}\alpha q_{l}^{2} &
   -\tfrac{1}{2}\alpha q_{l}^{2} &
   \tfrac{1}{\sqrt{2}}\beta q_{l}^{2} &
   0 & 0
\end{array}\right)\,,
\end{equation}
where the $s$-wave components are proportional to
\begin{equation}
F = -\frac{1}{4f^2}\left(k^0 + k^{\prime 0}\right),
\end{equation}
with $f = 93~\text{MeV}$ the pseudoscalar decay constant, and $k^0\,(k^{\prime 0})$ the energy of the incoming (outgoing) meson in the center-of-mass (CM) frame. The factor $F$ arises from the lowest-order Weinberg–Tomozawa term of the chiral SU(3) Lagrangian and provides the leading $s$-wave contribution to the meson–baryon interaction.

The off-diagonal elements proportional to $q_l^2$ account for $d$-wave transitions between $\Xi^\ast\olsi{K}$ or $\Omega\eta$ and the $\Xi\olsi{K}$ channels, with $l=\{\Xi^0K^-, \Xi^-\olsi{K}{}^0\}$.\footnote{In the $\Xi^\ast\olsi{K} \leftrightarrow \Xi\olsi{K}$ or $\Omega\eta \leftrightarrow \Xi\olsi{K}$ transitions, the $\Xi\olsi{K}$ states are in $d$-wave, but the $\Xi^\ast\olsi{K}$ are in $s$-wave, and thus the matrix elements carry only a relative $q_l^2$ factor. In the $\Xi\olsi{K} \leftrightarrow \Xi\olsi{K}$ transitions, both the initial and final states would be in $d$-wave and thus these matrix elements would carry a $q_l^4$ factor. However, at first approximation we neglect these latter transitions and can thus safely refer to the former as $d$-wave transitions.} The quantity
\begin{equation}
q_l = \frac{\lambda^{1/2}(s, M_{l}^2, m_{l}^2)}{2\sqrt{s}},
\end{equation}
denotes the on-shell relative momentum in channel $l$, where $\lambda$ is the Källén function and $M_l$ ($m_l$) are the baryon (meson) masses. The parameters $\alpha$ and $\beta$ are phenomenological coupling constants associated with the $\Xi^\ast\olsi{K} \!\leftrightarrow\! \Xi\olsi{K}$ and $\Omega\eta \!\leftrightarrow\! \Xi\olsi{K}$ transitions, respectively. They parametrize short-range dynamics not captured by the leading-order chiral amplitude and have dimensions of $[\text{MeV}^{-3}]$, consistent with the two powers of momentum in the $d$-wave vertices. As in Refs.~\cite{Pavao:2018xub,Ikeno:2020vqv}, the $d$-wave $\Xi\olsi{K} \to \Xi\olsi{K}$ interaction is neglected. In the present work, the parameters $\alpha$, and $\beta$ are determined by fitting the experimental properties of the $\Omega(2012)^{-}$ resonance, their values are presented in Sect.~\ref{sec:res}.

Since the $\Omega(2012)^{-}$ is an isosinglet ($I=0$) state, it couples only to the $I=0$ component of the meson–baryon pairs considered, which in the physical basis contain both $I=0$ and $I=1$ admixtures. In the absence of experimental information constraining the $I=1$ $\Xi^\ast\bar{K}\leftrightarrow \Xi\bar{K}$ $d$-wave transitions, their contribution is assumed to be negligible, as it is expected to be much smaller than the dominant $I=0$ amplitude associated with the $\Omega(2012)^-$. For the same reason, the $\pi^0\Omega^-$ channel, which is a pure $I=1$ state, is omitted from the calculation. Moreover, its threshold lies far below the $\Omega(2012)^-$ energy region, and its $s$-wave contribution is expected to be marginal.

With the transition potentials given in Eq.~\eqref{eq:Vmatrix}, the unitarized scattering matrix is obtained by solving the Bethe-Salpeter equation (BSE)
\newcommand{\regvspa}{\vphantom{\olsi{K}{}^0}}
\begin{equation}\label{eq:BSeq}
    T=(1-VG)^{-1}V\,,
\end{equation}
where $G=\text{diag}(G_{\Xi^{\ast0}K^- \regvspa}, G_{\Xi^{\ast-}\olsi{K}{}^0}, G_{\Omega^-\eta \regvspa }, G_{\Xi^0 K^- \regvspa}, G_{\Xi^-\olsi{K}{}^0}$) is a diagonal matrix, whose elements are given by the meson-baryon loop function as follows,
\begin{equation}
    G_j(s) = i \int \frac{\mathrm{d}^4 q}{(2\pi)^4} \frac{2M_j}{(P - q)^2 - M_j^2 + i\epsilon} \frac{1}{q^2 - m_j^2 + i\epsilon}\,.
\end{equation}
Since this loop function is ultraviolet (UV) divergent, we employ a hybrid regularization method, which consists in a subtraction scheme where the subtraction constant is parameterized in terms of an UV cutoff $\Lambda$. (For further details, see Refs.~\cite{Lin:2025mtz,Encarnacion:2025luc,Nieves:2024dcz}, in particular the detailed derivation in Appendix A of Ref.~\cite{Lin:2025mtz}). The resulting $G$ function reads
\begin{widetext}
\begin{eqnarray}
    G_j(s) & = &\frac{1}{4\pi^2} \frac{M_j}{m_j+M_j} \left
(m_j\ln\frac{m_j}{\Lambda + \sqrt{\Lambda^2+m_j^2}}+  M_j\ln\frac{M_j}{\Lambda + \sqrt{\Lambda^2+M_j^2}} \right)  +  \frac{2M_j}{16\pi^2} \frac{M_j-m_j}{M_j+m_j}\left( \frac{(M_j+m_j)^2}{s}-1\right) \ln \frac{M_j}{m_j}\nonumber\\
  &&+ \frac{M_j\sigma_j}{16\pi^2 s} \left\{ \ln\left( s - M_j^2 + m_j^2 + \sigma_j \right) - \ln\left( -s + M_j^2 - m_j^2 + \sigma_j \right) + \ln\left( s + M_j^2 - m_j^2 + \sigma_j \right) - \ln\left( -s - M_j^2 + m_j^2 + \sigma_j \right) \right\}\,,
\label{eq:G}
\end{eqnarray}
\end{widetext}
where $j=\{ \Xi^{\ast0}K^-, \Xi^{\ast-}\olsi{K}{}^0, \Omega^-\eta \}$, $\Lambda$ is the cutoff value, and $\sigma_j=2\sqrt{
s}p_j(s)$, and $p_j(s)=\lambda^{1/2}(s, m_j^2, M_j^2)/(2\sqrt{s})$. To provide flexibility in reproducing the available experimental data on the $\Omega(2012)^{-}$, the cutoff~$\Lambda$ is allowed to vary within a reasonable range around the expected $\rho$-meson mass.

For the $d$-wave channels $\Xi^0K^-$ and $\Xi^-\olsi{K}{}^0$, as in Refs.~\cite{Pavao:2018xub,Ikeno:2020vqv}, the $G$-functions become:
\begin{equation}\label{eq:Gd}
    \begin{aligned}
        G_{d-\text{wave}}^{(l)}(s)=&\int_{\abs{\vec{q}}<\Lambda^\prime} \frac{\text{d}^3q}{(2\pi)^3} \frac{(q/q_{l})^4}{2\omega_l(\vec{q}\,)} \frac{M_l}{E_{l}(\vec{q}\,)} \\
        &\times\frac{1}{\sqrt{s}-\omega_l(\vec{q}\,)-E_{l}(\vec{q}\,)+i\epsilon} \,,
    \end{aligned}
\end{equation}
where $\omega_l(\vec{q}\,)=\sqrt{\vec{q}^{\,2}+m_l^2}, E_l(\vec{q}\,)=\sqrt{\vec{q}^{\,2}+M_l^2}$.
The cutoff values $\Lambda$ and $\Lambda^\prime$ in Eqs.~\eqref{eq:G} and \eqref{eq:Gd} are not necessarily identical; however, for simplicity, we take them to be equal, as they are expected to differ only slightly.

The $\Omega(2012)^{-}$ lies about 10~MeV and 20~MeV below the $\Xi^{\ast0}K^-$ and $\Xi^{\ast-}\olsi{K}{}^0$ thresholds, respectively, so the loop functions of Eq.~\eqref{eq:G} for these channels are modified to account for the finite width of the $\Xi^{\ast}$:
\begin{equation}
\begin{aligned}
    \label{eq:G_tilde}
    \widehat{G}_j(s) =& \int_{M_j-6 \Gamma_{j}}^{M_j+6 \Gamma_{j}} \text{d} M \, \frac{1}{N_j} \operatorname{Im}\left[\frac{1}{M-M_j + i \Gamma_{j}/2}\right]
    \\
    &\times G_j\left(s, m_j, M\right)\,,
\end{aligned}
\end{equation}
with the normalization factor
\begin{equation}
    \label{eq:N}
    N_j=\int_{M_j-6 \Gamma_{j}}^{M_j+6 \Gamma_{j}} \text{d} M  \operatorname{Im}\left[\frac{1}{M-M_j+i \Gamma_j/2}\right]\,,
\end{equation}
where the decay widths $\Gamma_j$ of $\Xi^{\ast0}$ and $\Xi^{\ast-}$ are taken to be the same, $\Gamma_j=9.1~\text{MeV}$.
As a result, the $G$-matrix becomes $G=\text{diag}(\widehat{G}_{\Xi^{\ast0}K^- \regvspa}, \widehat{G}_{\Xi^{\ast-}\olsi{K}{}^0}, G_{\Omega^-\eta \regvspa}, G_{\Xi^0K^- \regvspa}, G_{\Xi^-\olsi{K}{}^0})$. To account for the bulk of the mass distribution, we find it safe to perform the integration over the range $M_{j}-6\Gamma_j$ to $M_{j}+6\Gamma_j$. We note that, due to the proximity of the $\Omega(2012)$ to the $\Xi^{\ast}\olsi{K}$ threshold, a detailed treatment of the $\Xi^{\ast}$ width—particularly its energy dependence—can be relevant in the description of the three-body decay mechanism $\Omega(2012)\to \Xi^{\ast}\olsi{K} \to \Xi \olsi{K} \pi$, as discussed in Refs.~\cite{Ikeno:2020vqv, Lu:2022puv}. In the present approach, however, a constant width is employed in the convolution. This approximation is justified by the relatively small value of the $\Xi^{\ast}$ width, which leads to a sharply peaked mass distribution, together with the limited near-threshold kinematic region probed in femtoscopy, where the energy dependence of the width is expected to be mild. 

Possible poles are searched for in the second Riemann sheet $G^{\text{II}}_j(s)$, which is given by
\begin{equation}
    \label{eq:G2}
    G^{\text{II}}_j(s) = G_j(s) + i \frac{2M_j}{4\pi\sqrt{s}} p_j(s)\,,
\end{equation}
for channels satisfying $\text{Re}[\sqrt{s}] > M_j + m_j$. In the vicinity of a pole, the scattering amplitudes behave as
\begin{equation}
    T_{ij} = \frac{g_ig_j}{\sqrt{s}-M_R+i\Gamma_R/2}\,,
\end{equation}
where $M_R$ and $\Gamma_R$ denote the pole mass and width, respectively.
The couplings $g_i$ of the resonance to the different channels are evaluated following Eqs.~(8)-(12) of Ref.~\cite{Ikeno:2020vqv}, both with and without including the width of the $\Xi^\ast$ baryon. 

\newcommand{\sqthi}{E_\text{th}^{(i)}}

\subsection{Scattering parameters}
The scattering length $a$ and effective range $r_0$ can be obtained from the relation between the $T$-matrix and the standard quantum-mechanical scattering amplitude,
\begin{equation}\label{eq:T-ERE-expansion}
    T = - \frac{8\pi \sqrt{s}}{2M} f^\text{QM} \simeq - \frac{8\pi \sqrt{s}}{2M} \frac{1}{\displaystyle -\frac{1}{a} + \frac{1}{2} r_0 p^2 - ip}\,,
\end{equation}
with $p$ the CM momentum of the particles. For each channel $i$, the scattering length $a_i$ and effective range $r_{0,i}$ can be determined from
\begin{subequations}\label{eq:ereparameters}
\begin{eqnarray}
    \label{eq:a}
    \frac{1}{a_i} &=& \left. \frac{8\pi \sqrt{s}}{2M_i} (T_{ii})^{-1} \right\rvert_{\sqthi}, \\[2mm]
    \label{eq:r}
    r_{0,i} &=& \left. \frac{1}{\mu_i} \frac{\partial}{\partial\sqrt{s}} \left[ \frac{-8\pi\sqrt{s}}{2M_i}(T_{ii})^{-1} + ip_i \right] \right\rvert_{\sqthi},
\end{eqnarray}
\end{subequations}
where $\mu_i = m_i M_i / (m_i + M_i)$ is the reduced mass of channel $i$, and $\sqthi$ is its threshold energy.
When the $\Xi^\ast$ width is included, the evaluation of $r_{0,i}$ requires replacing $p_i$ in Eq.~\eqref{eq:r} with 
\begin{equation}
 p_i \to -\frac{8\pi\sqrt{s}}{2M_i}\,\text{Im}[\widehat{G}_i] \,.   
\end{equation}
    
\subsection{Correlation function}
{

\begin{table*}[!htbp]
\centering
\caption{Values of the production weight for $\Xi^{\ast0}K^-, \Xi^{\ast-}\olsi{K}{}^0, \Omega^-\eta, \Xi^0K^-$, and $\Xi^- \olsi{K}{}^0$ CFs. The threshold mass of each channel, given in units of MeV, is shown in parentheses. Productions weights for the $\Xi^0 K^-$ and $\Xi^- \olsi{K}{}^0$ channels are shown for completeness, although these CFs are not computed in the present work.}
\setlength{\tabcolsep}{8pt}
{
\begin{tabular}{llccccccc}
\hline\hline
\multicolumn{2}{c}{channel$-j$} & $\Xi^{\ast0}K^-(2025)$ & $\Xi^{\ast-}\olsi{K}{}^0(2033)$ & $\Omega^-\eta(2220)$ & $\Xi^0K^-(1808)$ & $\Xi^- \olsi{K}{}^0(1819)$
\\
\hline
$[\Xi^{\ast0}K^- ~ \mathrm{CF}] $ & $ \omega_j$ & 
$1$ & $0.95\pm0.10$ & $0.089\pm0.011$ & $1.28\pm0.14$ & $1.23\pm0.13$ 
\\
\hline
$[\Xi^{\ast-}\olsi{K}{}^0 ~ \mathrm{CF}] $ & $ \omega_j$ & 
$1.04\pm0.11$ & $1$ & $0.096\pm0.011$ & $1.32\pm0.14$ & $1.27\pm0.13$ 
\\
\hline
$[\Omega^-\eta \quad\, \mathrm{CF}] $ & $ \omega_j$ & 
$3.04\pm0.36$ & $2.93\pm0.31$ & $1$ & $3.38\pm0.40$ & $3.26\pm0.39$
\\
\hline
$[\Xi^0K^- \quad\, \mathrm{CF}] $ & $ \omega_j$ & 
$0.17\pm0.02$ & $0.16\pm0.02$ & $0$ & $1$ & $0.93\pm0.09$
\\
\hline
$[\Xi^- \olsi{K}{}^0 \quad\, \mathrm{CF}] $ & $ \omega_j$ & 
$0.20\pm0.02$ & $0.18\pm0.02$ & $0$ & $1.06\pm0.11$ & $1$
\\
\hline\hline
\end{tabular}%

}%
\label{Tab:weight}
\end{table*}



%
The two-particle CF provides direct information on the final-state interactions (FSI) between hadrons produced in high-energy collisions. In the case of a meson-baryon system with coupled channels, the generalized KP formalism leads to the following expression for the CF of the measured channel $i$~\cite{Haidenbauer:2018jvl,Vidana:2023olz,Albaladejo:2024lam,Lin:2025mtz}: 
\begin{align}\label{eq:correlation}
\mathcal{C}_{i} (p_i) =& 1 + 4\pi\int_0^{\infty} \mathrm{d} r \, r^2 \,\, S_{12}(r)
\\
& \times\left\{ \sum_j \omega_j^{(i)} \left\lvert j_0(p_i r)\delta_{ij} + T_{ij}(\sqrt{s}) \widetilde{G}_j(r,s) \right\rvert^2 - j_0^2(p_ir)\right\}, \nonumber
\end{align}
where \(T_{ij}\) are the elements of the coupled-channel scattering matrix obtained in Eq.~\eqref{eq:BSeq}, and \(j_0(p_i r)\) is the spherical Bessel function of zeroth order. The possible relative weights between channels $\Xi^{\ast}\olsi{K}$-$\Omega\eta$ and $\Xi \olsi{K}$, due to their different orbital angular momentum, \textit{i.e.}, not the ones directly related to production, is however equal to one, as derived in detail in Appendix \ref{app:CF-JLS}. The source function \(S_{12}(\vec{r}\,)\) describes the spatial distribution of the particle-emitting source and is modeled as a spherically symmetric Gaussian:
\begin{equation}
    S_{12} (\vec{r} \, ) = \frac{1}{(\sqrt{4\pi})^3 R^3} \exp(-\frac{\vec{r}^{\,2}}{4R^2})\,,  
\end{equation}
where \(R\) denotes the effective source radius (source size), typically determined from $m_T$-scaling extrapolations. The propagator-like function $\widetilde{G}_j(r,s)$ in Eq.~\eqref{eq:correlation} is defined as
\begin{subequations}
\begin{align}
    \widetilde{G}_j(r, s) =& 2M_j \int_0^{\widetilde\Lambda} \frac{q^2 \mathrm{d} q}{2\pi^2}
    \frac{\Omega_j(q)+E_j(q)}{2 \Omega_j(q) E_j(q)}
    \\
    &\times\frac{j_0(q r)}{s-\left(\Omega_j(q)+E_j(q)\right)^2 + i\epsilon}\,, \nonumber
\end{align}
with $\Omega_j(q) = \sqrt{m_j^2 + \vec{q}^{\,2}}$ and $E_j(q) = \sqrt{M_j^2 + \vec{q}^{\,2}}$. For channels $j=\{\Xi^0 K^-,\Xi^- \olsi{K}{}^0\}$ it is necessary to modify the $\widetilde{G}(r,s)$ functions, to take into account that they are in $d$-wave, similarly as done for the $G(s)$ function [\textit{cf.} Eq.\,\eqref{eq:Gd}]. In particular, for these channels we have:
\begin{align}
    \widetilde{G}_j(r, s) =& 2M_j \int_0^{\widetilde\Lambda} \frac{q^2 \mathrm{d} q}{2\pi^2}
    \frac{\Omega_j(q)+E_j(q)}{2 \Omega_j(q) E_j(q)}
    \\
    &\times\frac{j_2(q r)}{s-\left(\Omega_j(q)+E_j(q)\right)^2 + i\epsilon} \left( \frac{q}{q_l(s)} \right)^2\,. \nonumber
\end{align}
Note that only a factor $q^2$ is introduced. This is because only the amplitude $T$ is factored out---indeed, the factor $j_2(qr)$ already behaves as $q^2$ near the $\Xi \olsi{K}$ thresholds.
\end{subequations}
The resulting correlation functions are found to be only weakly dependent on the value of $\widetilde\Lambda$, owing to the effective regulator provided by the combined action of the Bessel function $j_0(qr)$ and relativistic kinematics. 
Therefore, the $\widetilde\Lambda$ cutoff value employed to truncate the $\widetilde{G}_j$ integral is taken to be the same as in Eqs.~\eqref{eq:G} and \eqref{eq:Gd}, following Refs.\,\cite{Vidana:2023olz,Feijoo:2024bvn}.  Finally, the production weights $\omega_{j}^{(i)}$ entering Eq.\,\eqref{eq:correlation} quantify the relative contributions of different coupled channels and are summarized in Table~\ref{Tab:weight}. These values are computed following the VLC method described in Appendix~A of Ref.\,\cite{Encarnacion:2024jge}, under the assumption of high-multiplicity events in $p-p$ collisions. Accordingly, the parameters of the $\gamma_s$CSM model are taken from Table III of Ref.~\cite{Encarnacion:2024jge}, consistently with the multiplicity-dependent analysis of Ref.~\cite{Vovchenko:2019kes}. The parameters of the blast-wave (CBW) model used to obtain the single-particle momentum distributions are also taken from Table III of Ref.~\cite{Encarnacion:2024jge}. Since they are extracted from fits to experimental momentum spectra for different particle species, these values can be applied more generally to hadrons containing light and strange quarks. The cutoff used to integrate the center-of-mass momentum distributions is set to $k_{\rm max}=500$ MeV.
\section{Results}\label{sec:res}
As discussed in Sect.~\ref{sec:Form}, our model depends on the parameters $\Lambda$, $\alpha$, and $\beta$, which are determined by fitting to the measured properties of the $\Omega(2012)^{-}$ resonance. The relevant experimental values, taken from the RPP~\cite{ParticleDataGroup:2024cfk} are 
\begin{subequations}\label{eq:exp_data}
\begin{eqnarray}
    M_{\Omega(2012)}^{\text{exp}} &=& (2012.5 \pm 0.6) \, \text{MeV}\,, 
    \\
    \Gamma_{\Omega(2012)}^{\text{exp}} &=& (6.4^{+3.0}_{-2.6}) \, \text{MeV}\,,
    \\
    \mathcal{R}^{\Xi\olsi{K}\pi}_{\Xi\olsi{K}}&=&\frac{\Gamma(\Xi\olsi{K}\pi)}{\Gamma(\Xi\olsi{K})} = 0.99 \pm 0.27\,.    
\end{eqnarray}
\end{subequations}
By incorporating the spectral mass distribution of the $\Xi^\ast$ into the loop function $G_{\Xi^\ast K}$ of Eq.~\eqref{eq:G_tilde}, the sharp threshold structure of the zero-width limit is smoothed out, which complicates a proper definition of the Riemann-sheet structure. Therefore, the $\Omega(2012)^-$ mass is identified with the value of $\sqrt{s}$ corresponding to the maximum of $\abs{T_{\Xi^{\ast0}K^-\to\Xi^{\ast0}K^-}}^2$, whereas the width is determined from the full width at half maximum of the same quantity.
The branching fraction is calculated as the ratio between the partial decay widths $\Gamma(\Xi\olsi{K}\pi)$ and $\Gamma(\Xi\olsi{K})$, which are evaluated according to Eq.~(19) of Ref.~\cite{Ikeno:2020vqv} and Eq.~(15) of Ref.~\cite{Lu:2020ste}, respectively, using the inputs derived within the present framework.


Besides employing a different regularization prescription, it is worth emphasizing that the inclusion of the branching fraction $\mathcal{R}^{\Xi\olsi{K}\pi}_{\Xi\olsi{K}}$ in the fitting procedure constitutes a key improvement with respect to Refs.~\cite{Pavao:2018xub,Ikeno:2020vqv}. The best-fit values are:
\begin{subequations}\label{eq:fit_para}
    \begin{eqnarray}
        \Lambda &=& (814 \pm 1)\,\text{MeV}\,,\\
        \alpha &=& (3.62 \pm 0.17)\times10^{-8}\, \text{MeV}^{-3}\,,\\
        \beta &=& (1.07 \pm 0.10)\times10^{-8}\, \text{MeV}^{-3}\,.
    \end{eqnarray}
\end{subequations}
The fitted values of $\alpha$ and $\beta$ are consistent with those reported in Ref.~\cite{Pavao:2018xub}, and their magnitudes are of the same order as those obtained in Ref.~\cite{Valderrama:2018bmv}. There are differences with respect to Ref.\,\cite{Shen:2025xcq}, although the parameters are of similar scale. These differences can be ascribed to the different regularization method and to the different fitting procedure.

It is also worth noting that the covariance matrix $r_{ij}$ obtained in the fit exhibits an almost uncorrelated pattern, with diagonal dominance ($r_{ii}\simeq 1$) and small off-diagonal elements: $r_{\Lambda\alpha}=0.07$, $r_{\Lambda\beta}=0.03$, and $r_{\alpha\beta}=0.07$. This indicates that the fitted parameters are largely independent. 

Using the parameters values in Eq.~\eqref{eq:fit_para}, we obtain the following results when the $\Xi^\ast$ width is taken into account:
\begin{subequations}\label{eq:fit_results}
\begin{eqnarray}
    M_{\Omega(2012)}^{\text{fit}} &=& (2012.53 \pm 0.73)\,\text{MeV}\,,\\
    \Gamma_{\Omega(2012)}^{\text{fit}} &=& (4.05 \pm 0.13)\,\text{MeV}\,,\\
    \left[\mathcal{R}^{\Xi\olsi{K}\pi}_{\Xi\olsi{K}}\right]^{\text{fit}} &=& 0.95 \pm 0.10\,.
\end{eqnarray}
\end{subequations}
For completeness, we also present the results obtained when the $\Xi^\ast$ width is not taken into account in the loop function $G_{\Xi^\ast\olsi{K}}$ and taking the values of the parameters $\Lambda$, $\alpha$, and $\beta$ extracted from the fit. In this case, the loop function $G_{\Xi^\ast\olsi{K}}$ given in Eq.~\eqref{eq:G} is employed, and the pole position is searched by solving $\det(1 - VG) = 0$ on the second Riemann sheet. The corresponding $\Omega(2012)$ pole position obtained is:
\begin{equation}
\begin{aligned}
M_{\Omega(2012)} -i\frac{\Gamma_{\Omega(2012)}}{2}= 
(2013.32 - i\,1.15)\,\text{MeV}\,\,.
\end{aligned}\label{eq:pole_locations}
\end{equation}
Therefore, it can be appreciated that, when the $\Xi^\ast$ width is included, the resulting mass and width of the $\Omega(2012)$ state [Eqs.~\eqref{eq:fit_results}] lie well within the experimental ranges listed in Eq.~\eqref{eq:exp_data}. In contrast, when the $\Xi^\ast$ width is neglected, the obtained width is $2.3\,\text{MeV}$, which is slightly below the experimental interval but still in reasonable agreement with the measured value.

\begin{table*}[!htbp]
\centering
\caption{Couplings of $\Omega(2012)$ to each channel $g_i$, wave functions at the origin $g_iG_i$ [in MeV], and compositeness magnitude $-g_i^2\tfrac{\partial G_i}{\partial\sqrt{s}}$ obtained from the pole for the case excluding the $\Xi^\ast$ width. Couplings $\tilde{g}_i$ obtained from the pole for the case including the $\Xi^\ast$ width.}
\setlength{\tabcolsep}{9pt}
\begin{tabular}{cccccc}
\hline
\hline
 & $\Xi^{* 0} K^{-}$ & $\Xi^{*-} \olsi{K}{}^0$ & $\Omega^{-} \eta$ & $\Xi^0 K^-$ & $\Xi^-\olsi{K}{}^0$ \\
\hline
$\hphantom{+}g_i$ & $1.16+i0.01$ & $1.21+i0.01$ & $-2.69-i0.00$ & $-0.17+i0.00$ & $0.16+i0.00$
\\
\hline
$\hphantom{+}\tilde{g}_i$ & $1.16-i0.00$ & $1.21-i0.04$ & $-2.69 + i0.09$ & $-0.17+i0.01$ & $0.16-i0.01$
\\
\hline
$\hphantom{+}g_iG_i$ & $-25.13+i0.03$ & $-24.34+i0.01$ & $20.08 - i0.11$ & $\cdots$ & $\cdots$
\\
\hline
$-g_i^2\tfrac{\partial G_i}{\partial \sqrt{s}}$ & $0.28-i0.01$ & $0.23-i0.00$ & $0.27 - i0.00$ & $\cdots$ & $\cdots$
\\
\hline
\hline
\end{tabular}%
\label{tab:gi}
\end{table*}
\begin{figure*}[!htbp]
\begin{center}
\includegraphics[width=1.0\textwidth]{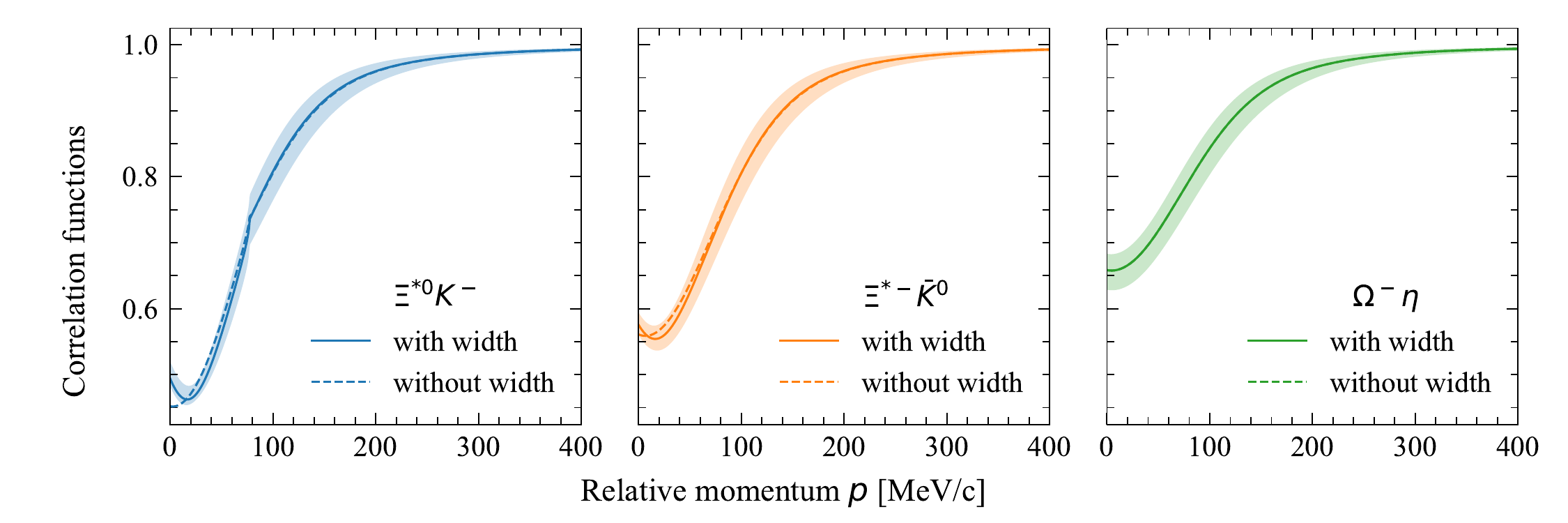}
\end{center}
\vspace{-0.7cm}
\caption{Correlation functions of $\Xi^{\ast0}K^-, \Xi^{\ast-}\olsi{K}{}^0$, and $\Omega^-\eta$ channels with source size $R=1.2\,\text{fm}$. The solid and dashed lines correspond to the cases with and without the $\Xi^\ast$, respectively. The 68\% CL color-shaded bands are estimated from the uncertainties in the production weights, the parameters $\Lambda, \alpha, \beta$, and the 10\% uncertainty in the source size.
\label{Fig:cor_func_nowidth}}
\end{figure*}
%
%

Table~\ref{tab:gi} summarizes the results for the couplings of the $\Omega(2012)$ resonance to each channel, $g_i$, for both cases considered. The table also lists the corresponding wave functions at the origin, $g_i G_i$, and the compositeness magnitude, $-g_i^2\partial G_i/\partial \sqrt{s}$, for the $\Xi^{\ast 0}K^-$, $\Xi^{\ast-}\olsi{K}{}^0$, and $\Omega^-\eta$ channels obtained without including the $\Xi^\ast$ width. The compositeness magnitude, $-g_i^2\partial G_i/\partial \sqrt{s}$, quantifies the probability of finding a given channel as a bound component of the state~\cite{Gamermann:2009uq,Sekihara:2014kya,Kamiya:2015aea,Albaladejo:2022sux}. For open channels, however, it corresponds to the integral of the squared wave function, defined with an appropriate phase prescription~\cite{Aceti:2012dd}. It is worth noting that both the couplings and the loop functions $G_i$ become complex when some channels are open. In such cases, the compositeness magnitude can no longer be interpreted as a strict probability; nevertheless, it remains a meaningful quantity, providing a quantitative estimate of the relative weight of each component in the total wave function. Alternatively, in such a situation, the magnitude $g_i G_i$ conveys similar information. 

From the last two rows of Table~\ref{tab:gi}, one can conclude that the $\Omega(2012)$ resonance exhibits a predominantly molecular nature, with a total compositeness of about $78\%$.
The similarity of the relative contributions obtained for both $g_i G_i$ and $-g_i^2\partial G_i/\partial \sqrt{s}$ across the considered channels highlights their comparable roles in the $\Omega(2012)$ dynamics.

\begin{table}[t]
\centering
\caption{Scattering length $a$ and effective range $r_0$ for the cases without and with $\Xi^\ast$ width taken into account. [in fm]}
\setlength{\tabcolsep}{7pt}
\begin{tabular}{cccccc}
\hline
\hline
 & $\Xi^{* 0} K^{-}$ & $\Xi^{*-} \olsi{K}{}^0$ & $\Omega^{-} \eta$ \\
\hline
\hline
\multicolumn{4}{c}{Without $\Xi^\ast$ width }\\
\hline
$a$ & $\hphantom{+}1.06-i0.06$ & $\hphantom{+}0.74-i0.12$ & $0.45-i0.04$
\\
\hline
$r_0$ & $-3.67+i0.11$ & $-1.77-i0.77$ & $0.05-i0.00$
\\
\hline
\hline
\multicolumn{4}{c}{With $\Xi^\ast$ width}\\
\hline
$a$ & $1.14-i0.37$ & $0.80-i0.26$ & $0.45 - i0.04$
\\
\hline
$r_0$ & $-12.87-i0.17$ & $-11.36-i0.65$ & $0.05 - i0.00$
\\
\hline
\hline
\end{tabular}%
\label{tab:sca_para}
\end{table}

Next, we present in Table~\ref{tab:sca_para} the results for the scattering parameters obtained both with and without including the finite width of the $\Xi^\ast$ baryon. As seen in the table, and taking into account the sign convention adopted in Eq.~\eqref{eq:T-ERE-expansion}, the positive and large values of the $\Xi^\ast\bar{K}$ scattering lengths reflect the strong attractive interaction that dynamically generates the $\Omega(2012)$ resonance. Although the inclusion of the $\Xi^\ast$ width does not qualitatively alter the scattering lengths, it leads to slightly larger values compared to the zero-width case. On the other hand, the effective ranges $r_0$ for the $\Xi^{\ast0}K^-$ and $\Xi^{\ast-}\olsi{K}{}^0$ channels exhibit a more pronounced sensitivity to the $\Xi^\ast$ width, reflecting its impact on the near-threshold dynamics.

Finally, we present the first theoretical predictions for the correlation functions in the $S=-3$, $Q=-1$ sector, obtained within the present chiral unitary approach.
Fig.~\ref{Fig:cor_func_nowidth} shows the CFs for the $\Xi^{\ast0}K^-$, $\Xi^{\ast-}\olsi{K}{}^0$, and $\Omega^-\eta$ channels, where the presence of the $\Omega(2012)$ resonance may influence their behavior or leave observable imprints. The results correspond to a source size of $R=1.2$~fm, with dashed and solid lines denoting calculations excluding and including the finite width of the $\Xi^\ast$ baryon, respectively. The 68\% confidence-level (CL) bands are obtained via a Monte Carlo sampling procedure that propagates the uncertainties from the production weights listed in Table~\ref{Tab:weight}, the parameters $\Lambda$, $\alpha$, and $\beta$ given in Eq.~\eqref{eq:fit_para}, as well as a 10\% uncertainty in the source size.

For completeness, we note that although the $\Xi \bar K$ channels are included in the present framework and exhibit a visible imprint of the $\Omega(2012)$, the corresponding correlation functions are not shown. In the present approach, the $\Xi \bar K$ interaction enters only through $d$-wave contributions, while other potentially relevant partial waves, particularly the $s$- and $p$-wave components expected to dominate the femtoscopic momentum region, are not included\footnote{Near-threshold $\Xi\bar K$ CFs have already been studied within a chiral Lagrangian approach up to next-to-leading order in Ref.~\cite{Feijoo:2024qgq}.}. In addition, the convolution with the source function tends to wash out the characteristic shape of the correlation function, making the resulting correlation functions not suitable for quantitative comparison with experiment.

Focusing first on the $\Xi^*\bar{K}$ channels, shown in the left and central panels of Fig.~\ref{Fig:cor_func_nowidth}, the main feature is the presence of the $\Omega(2012)$ resonance, located slightly below the corresponding thresholds---by about $10~\text{MeV}$ for the $\Xi^{\ast0}K^-$ channel and $20~\text{MeV}$ for the $\Xi^{\ast-}\bar{K}^0$ one---which appears as a valley-like structure at threshold (behavior discussed in \cite{Liu:2023uly}). This behavior is observed in both calculations, with and without inclusion of the $\Xi^*$ width, yielding nearly identical correlation functions except for a mild difference in curvature near threshold, attributable to the small shift in the pole position shown in Eqs.~\eqref{eq:fit_results} and \eqref{eq:pole_locations}. A subtle shoulder is also seen in the $\Xi^{\ast0}K^-$ channel around a relative momentum of $80~\text{MeV}/c$, corresponding to the opening of the $\Xi^{\ast-}\bar{K}^0$ channel. In contrast, the $\eta\Omega^-$ CF curves exhibit no visible difference between the cases with and without inclusion of the $\Xi^*$ width. This is due to two reasons. First, the $\eta\Omega^-$ threshold lies far above the energy region of interest, preventing any influence from the $\Xi^*$ spectral distribution. Second, the small variation in the pole position has no impact on this channel because of the remoteness of its threshold. 
From these observations, it follows that only the $\Xi^*\bar{K}$ correlation functions carry discernible traces of the $\Omega(2012)$ resonance, with the $\Xi^{\ast0}K^-$ CF being the most sensitive owing to its threshold proximity to the resonance position.

To conclude, Fig.~\ref{Fig:cor_func_different} shows the contributions of the different coupled channels to the $\Xi^{\ast0}K^-$ CF in the case where the $\Xi^\ast$ width is neglected for simplicity. The CF is first evaluated including only the $\Xi^{\ast0}K^-$ channel in Eq.~\eqref{eq:correlation}, and the remaining channels are progressively added.

As seen in the figure, all channels contribute to the total $C_{\Xi^{\ast0}K^-}$ CF to varying degrees, depending on their interplay and on the values of $\omega_j^{(\Xi^{\ast0}K^-)}$. The latter are slightly larger for the lighter channels, whereas the nearly vanishing weight of the $\eta\Omega^-$ channel suppresses its inelastic contribution to the total $C_{\Xi^{\ast0}K^-}$ CF (see the first row of Table~\ref{Tab:weight}). The elastic transition accounts for about 78\% of the total $\Xi^{\ast0}K^-$ CF, whereas roughly 18\% originates from the inelastic $\Xi^{\ast-}\bar{K}{}^0$ contribution, leaving about 4\% associated with the $\Omega\eta$ and $\Xi\bar{K}$ inelastic channels. 

Although the $\eta\Omega^-$ channel contributes only marginally to the $\Xi^{\ast0}K^-$ CF, it plays a crucial role in the amplitudes, entering Eq.~\eqref{eq:correlation}, through the unitarization procedure when solving BSE. The relevance of this channel is evidenced not only by the corresponding coupling constants and compositeness values listed in Table~\ref{tab:gi}, but also by the distinctive features of the Weinberg–Tomozawa interaction in this sector. In fact, the elastic $\Xi^{\ast}\bar{K}$ and $\eta\Omega$ components of the kernel in the $I=0$ sector vanish, indicating that the $\Xi^{\ast}\bar{K}$ system cannot bind by itself. It is the coupling to the $\eta\Omega$ channel that ultimately generates the bound state. Consequently, both channels play an inseparable role in the dynamical generation of the $\Omega(2012)$ resonance.
\begin{figure}[t]
\begin{center}
\includegraphics[width=0.45\textwidth]{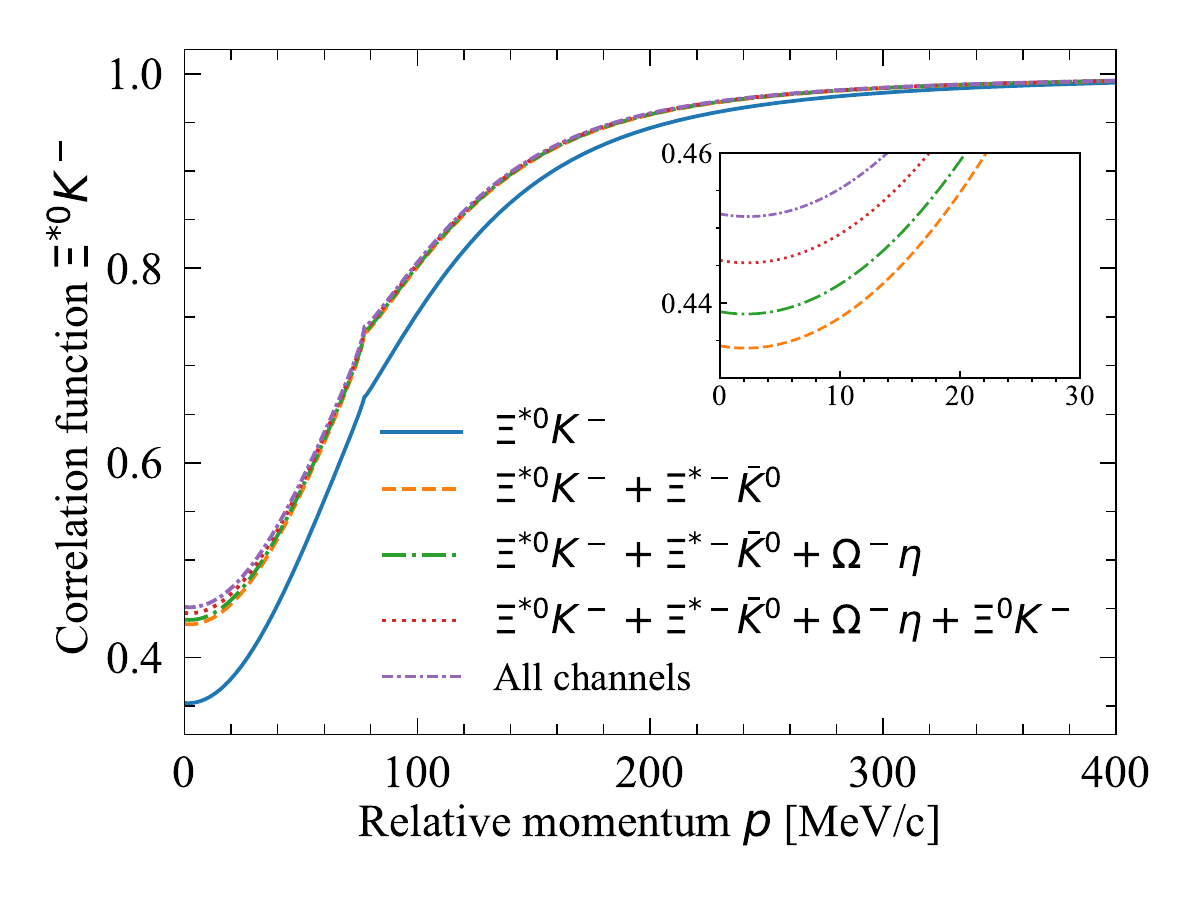}
\end{center}
\vspace{-0.7cm}
\caption{The contributions of different coupled channels to $\Xi^{\ast0}K^-$ CF.
\label{Fig:cor_func_different}}
\end{figure}

\section{Conclusions}

We have revisited the $\olsi{K}\Xi(1530)$ interaction in the $S=-3$, $Q=-1$ sector within a coupled-channel chiral unitary framework consistent with the properties of the $\Omega(2012)^-$. 
By fitting the model parameters to reproduce its measured mass, width, and branching ratio, we achieved a quantitative description of the state. 
The results confirm that the $\Omega(2012)^-$ is predominantly a dynamically generated $\Xi^\ast\bar{K}$--$\Omega\eta$ molecular state with a total compositeness of about $80\%$. 
The large $\Xi^\ast\olsi{K}$ scattering lengths further support a strong attractive interaction driving its formation. 
The predicted femtoscopic correlation functions for $\Xi^{\ast0}K^-$ and $\Xi^{\ast-}\olsi{K}{}^0$ pairs reveal clear threshold features correlated with the $\Omega(2012)$ pole, identifying the $\Xi^{\ast0}K^-$ CF as the most sensitive observable of its structure. 
Our findings highlight the crucial role of the $\eta\Omega$ channel in the coupled-channel dynamics and provide quantitative benchmarks for forthcoming femtoscopic measurements at the LHC, offering a direct test of the dominant molecular nature of the $\Omega(2012)^-$.

\section*{Acknowledgments}
The authors are very grateful to E. Oset for the fruitful discussions and his careful reading of the manuscript. This work is supported by the Spanish Ministerio de Ciencia e Innovaci\'on (MICINN) under contracts PID2020-112777GB-I00, PID2023-147458NB-C21 and CEX2023-001292-S; by Generalitat Valenciana under contracts PROMETEO/2020/023 and  CIPROM/2023/59. M.\,A. acknowledges financial support through \guillemotleft{}Ramón y Cajal\guillemotright{} program by MICINN Grant No.\,RYC2022-038524-I, and \guillemotleft{}Atracción de Talento\guillemotright{} program by CSIC PIE 20245AT019. J.X. Lin acknowledges the support of the China Scholarship Council. M. A. and A. F. thank the warm support of the ACVJLI.%

\bibliographystyle{apsrev4-1}
\bibliography{refs}

\onecolumngrid

\appendix

\newcommand{\vpp}{\vec{p}{}'}
\newcommand{\vp}{\vec{p}}
\newcommand{\vk}{\vec{k}}
\renewcommand{\vr}{\vec{r}}
\newcommand{\up}{\hat{p}}
\newcommand{\uk}{\hat{k}}
\newcommand{\ur}{\hat{r}}

\section{\boldmath Structure of correlation functions for arbitrary $JLS$}\label{app:CF-JLS}


In this Appendix we make a derivation of the structure of the CF when several channels with different spins and angular momenta are involved. For simplicity, the derivation is carried out for non-relativistic states and kinematics. For a two-particle system of the species $\alpha$ having a threshold $E_{\alpha}$, with CoM relative on-shell momentum $\vk$ and energy $E=E_\alpha + \vk^2/(2\mu_\alpha)$, and with total spin and third component of spin $S$ and $M_S$, we define free and interacting states as $\ket*{\vk;S,M_S;\alpha}$ and $\ket*{\Psi_{\vk};S,M_S;\alpha}$, respectively. The $T$-matrix operator is denoted with $\widehat{T}(E)$, and the free propagator operator is denoted as $\widehat{G}_0(E)$. The LS equation reads:
\begin{equation}\label{eq:LS}
\ket*{\Psi_{\vk};S,M_S;\alpha} = \ket*{\vk;S,M_S;\alpha} + \widehat{G}_0(E) \widehat{T}(E) \ket*{\vk;S,M_S;\alpha}\,.
\end{equation}
A resolution of the identity is:
\begin{equation}
    \mathbb{I} = \sum_{S',M'_S,\alpha'} \int d^3\vp \dyad*{\vp;S',M'_S;\alpha'}\,,
\end{equation}
so that $\widehat{G}_0(E)$ can be writen as:
\begin{equation}\label{eq:G0}
\widehat{G}_0(E) = \sum_{S',M'_S,\alpha'} \int d^3\vp \frac{\dyad*{\vp;S',M'_S;\alpha'}}{\displaystyle E-E_{\alpha'} -\frac{\vp^2}{2\mu_{\alpha'}}}\,.
\end{equation}
One of the components of the wave function in coordinate space is:\footnote{We define $\mathcal{N} = (2\pi)^\frac{3}{2}$. The normalizations are taken from Ref.\,\cite{Albaladejo:2024lam}.}
\begin{align}
\mathcal{N}^{-1} \, \Psi_{S,M_S;\alpha}^{(S',M'_S;\alpha')}(\vr;\vk) \equiv &
\braket*{\vr;S',M'_S;\alpha'}{\Psi_{\vk};S,M_S;\alpha} = 
\braket*{\vr;S',M'_S;\alpha'}{\vk;S,M_S;\alpha} \\
& + \sum_{S'',M''_S,\alpha''} \int d^3\vp \frac{\braket*{\vr;S',M'_S;\alpha'}{\vp;S'',M''_S;\alpha''}\mel*{\vp;S'',M''_S;\alpha''}{\widehat{T}}{\vk;S,M_S;\alpha}}{\displaystyle E-E_{\alpha''} -\frac{\vp^2}{2\mu_{\alpha''}}} \nonumber\,,
\end{align}
where we have used Eqs.\,\eqref{eq:LS} and \eqref{eq:G0}. Now, we use the plain-wave expression:
\begin{equation}
\mathcal{N}\,\braket*{\vr;S',M'_S;\alpha'}{\vk;S,M_S;\alpha} = \delta_{S,S'}\,\delta_{M_S,M'_S}\,\delta_{\alpha,\alpha'}\,e^{i\vk\cdot\vr} \,,
\end{equation}
to obtain:
\begin{equation}\label{eq:Psi_intermediate}
\Psi_{S,M_S;\alpha}^{(S',M'_S;\alpha')}(\vr;\vk) = \delta_{S,S'}\,\delta_{M_S,M'_S}\,\delta_{\alpha,\alpha'} \, e^{i\vk\cdot\vr} + \int d^3\vp \, e^{i\vp\cdot\vr}\,\frac{\mel*{\vp;S',M'_S;\alpha'}{\widehat{T}}{\vk;S,M_S;\alpha}}{\displaystyle E-E_{\alpha'} -\frac{\vp^2}{2\mu_{\alpha'}}}\,,
\end{equation}
Now we decompose the $\ket*{\vk;S,M_S;\alpha}$ states into $\ket*{k;JM;LS}$ states, this is to say, states with total angular momentum $J$, third component $M$, and orbital angular momentum $L$ (as well as spin $S$, as previously defined) as:
\begin{equation}
\ket*{\vk;S,M_S;\alpha} = \ket*{k,\Omega_{\uk};S,M_S;\alpha} = \!\! \sum_{L,M_L} \!\! Y_L^{M_L}(\Omega_{\uk})^{\ast} \ket{k;L,M_L;S,M_S;\alpha} = \!\!\!\! \sum_{L,M_L,J,M} \!\!\!\! Y_L^{M_L}(\Omega_{\uk})^{\ast} \braket{L,M_L;S,M_S}{J,M}\ket{k;J,M;L,S;\alpha}\,.
\end{equation}
We denote by $\Omega_{\hat{q}}$ the angles that define the direction of any unitary vector $\hat{q}$. For the amplitude, one has:\footnote{Note that the factor $4\pi$ is such that, for the case $S=S'=0$, one recovers the usual formula $\mel*{\vp}{\widehat{T}}{\vk} = \sum_{L}(2L+1)P_L(\up \cdot \uk) t_L(p,k)$\,.}
\begin{equation}
\mel*{\vp;S',M'_S;\alpha'}{\widehat{T}}{\vk;S,M_S;\alpha} = 4\pi \!\!\!\!\!\!\!\!\! \sum_{J,M,L,M_L,L',M'_L} \!\!\!\!\!\!\!\!\! 
Y_{L }^{M_{L }}(\Omega_{\uk})^{\ast}
Y_{L'}^{M_{L'}}(\Omega_{\up}) \braket*{L,M_L;S,M_S}{J,M}\!\!\braket*{L',M'_L;S',M'_S}{J,M} t^J_{LS\alpha;L'S'\alpha'}(k,p)\,.
\end{equation}
Inserting this decomposition and the one for $e^{i\vk\cdot\vr}$ in terms of the spherical harmonics of $\Omega_{\hat{k}}$ and $\Omega_{\hat{r}}$, one can write the wave-function component in Eq.\,\eqref{eq:Psi_intermediate} as:
\begin{subequations}\begin{equation}\label{eq:Psi_YLML_decomposition}
\Psi_{S,M_S;\alpha}^{(S',M'_S;\alpha')}(\vr;\vk) = 4\pi \!\!\!\!\!\! \sum_{L,M_L,L',M'_L} \!\!\!\!\!\! i^L Y_{L}^{M_L}(\Omega_{\uk})^{\ast} Y_{L'}^{M_L'}(\Omega_{\ur}) \, h_{S,M_S;L,M_L;\alpha}^{(S',M'_S;L',M'_L;\alpha')}(r,k)\,,
\end{equation}
with:\footnote{Notice that the phase $i^{L'-L}$ is not relevant for the CFs. If $L'=L$, then $i^{L'-L} = 1$, and if $L'\neq L$, then the first term of $h$ is zero. Hence, $i^{L'-L}$ becomes a global phase that vanishes when the absolute value is taken.}
\begin{align}\label{eq:h_A}
h_{S,M_S;L,M_L;\alpha}^{(S',M'_S;L',M'_L;\alpha')}(r,k) & = \delta_{\alpha,\alpha'}\delta_{L,L'}\delta_{M_L,M'_L}\delta_{S,S'}\delta_{M_S,M'_S} \, j_L(kr) \nonumber \\
& + i^{L'-L} \sum_{J,M} \braket*{L',M'_L;S',M'_S}{J,M} \! \braket{L,M_L;S,M_S}{J,M} I^J_{LS\alpha;L'S'\alpha'}(r,k)\,,\\
I^J_{LS\alpha;L'S'\alpha'}(r,k) & = 4\pi \int p^2 dp \frac{t^J_{LS\alpha;L'S'\alpha'}(k,p) \, j_{L'}(pr) }{\displaystyle E-E_{\alpha'} -\frac{p^2}{2\mu_{\alpha'}} }\,.
\end{align}
\end{subequations}
Next, using in Eq.\,\eqref{eq:h_A} the following identity:
\begin{equation}
\sum_{J,M} \braket*{L,M_L;S,M_S}{J,M} \!\! \braket*{L,M'_L;S,M'_S}{J,M} = \delta_{M_L,M'_L}\delta_{M_S,M'_S}\,,
\end{equation}
we can write:
\begin{equation}\label{eq:hSMSLMLtoJLS}
h_{S,M_S;L,M_L;\alpha}^{(S',M'_S;L',M'_L;\alpha')}(r,k) = \sum_{J,M} \braket*{L,M_L;S,M_S}{J,M} \!\! \braket*{L',M'_L;S',M'_S}{J,M} h^{J}_{LS\alpha;L'S'\alpha'}(r,k)\,,
\end{equation}
with:
\begin{equation}\label{eq:hJLS}
    h^{J}_{LS\alpha;L'S'\alpha'}(r,k) = \delta_{LL'}\delta_{SS'}\delta_{\alpha\alpha'} j_L(kr) \epsilon_{JLS}\epsilon_{J L'S'} + i^{L'-L} I^J_{LS\alpha,L'S'\alpha'}(r,k)\,,
\end{equation}
where the \textit{triangular coefficient}, $\epsilon_{JLS}$, is such that it is one when $J$ can be achieved with the angular momentum addition of $L$ and $S$, and zero elsewhere. 

The CF can be written as:
\begin{equation}\label{eq:CF_versionA}
    C^{\alpha}(k) =\int d^3\vec{r} \, S(\vec{r}) \, \frac{1}{4\pi}\int \!\! d\Omega_{\uk} \, \frac{1}{(2s_1+1)(2s_2+1)} \sum_{S,M_S}\sum_{S',M'_S,\alpha'} \left\lvert \Psi_{S,M_S;\alpha}^{(S',M'_S;\alpha')}(\vr;\vk) \right\rvert^2\,.
\end{equation}
Above, $\displaystyle\frac{1}{4\pi}\int d\Omega_{\uk}$ represents the angular average over all possible momenta $\vec{k}$, whereas $$\displaystyle\frac{1}{(2s_1+1)(2s_2+1)} \sum_{S,M_S}\sum_{S',M'_S,\alpha'}$$ represents the sum and average over the possible intermediate and final state spins, respectively, including a sum over all possible particle species $\alpha'$. Here, $s_1$ and $s_2$ respectively denote the spin of particles $1$ and $2$ in the two-body system measured in the CF. 

The sum over all possible intermediate states can also be understood as follows. Define $\widehat{P}(\vr)$ as:
\begin{equation}
    \widehat{P}(\vr) = \sum_{S',M'_S,\alpha'} \dyad*{\vr;S',M'_S,\alpha'}\,.
\end{equation}
Then, the wave-function squared absolute value is the mean value of this operator, which in turn is the sum of the squared absolute values of the wave-function components,
\begin{equation}
    \left\lvert \psi_{\vk;S,M_S;\alpha}(\vr) \right\rvert^2 = \mel*{\psi_{\vk};S,M_S;\alpha}{\widehat{P}(\vr)}{\psi_{\vk};S,M_S;\alpha} = \sum_{S',M'_S,\alpha'} \left\lvert \Psi_{S,M_S;\alpha}^{(S',M'_S;\alpha')}(\vr;\vk) \right\rvert^2\,.
\end{equation}
Alternatively, one can regard the source as an operator $\widehat{S}$:
\begin{equation}
    \widehat{S} = \sum_{S',M'_S,\alpha'} \int d^3\vr\,S(\vr)\,\dyad*{\vr;S',M'_S,\alpha'} = \int d^3\vr \, S(\vr)\widehat{P}(\vr)\,,
\end{equation}
and
\begin{equation}
    C^{\alpha}(k) = \frac{1}{4\pi}\int \!\! d\Omega_{\uk} \, \frac{1}{(2s_1+1)(2s_2+1)} \sum_{S,M_S} \mel*{\psi_{\vk};S,M_S,\alpha}{\widehat{S}}{\psi_{\vk};S,M_S,\alpha}\,.
\end{equation}
Note that we have assumed that the production weights are the same for each channel $\alpha$ in order to have a more compact notation. Nonetheless, the formulae can be easily generalized, \textit{i.e.} by replacing $S(\vr) \rightarrow \omega_{\alpha'}S_{\alpha'}(\vr)$.

We now return to Eq.\,\eqref{eq:CF_versionA}, inserting Eq.\,\eqref{eq:Psi_YLML_decomposition}. We assume a spherically symmetric source $S(\vec{r})$, so that the angular integrations are straightforward:
\begin{align}
\frac{1}{4\pi} & \int d\Omega_{\uk} \int d\Omega_{\ur} \hspace{-20pt} \sum_{%
{\scriptsize\begin{array}{c}
L\,,L'\,,\olsi{L}\,,\olsi{L}{}'\\
M_L\,,M'_L\,,\olsi{M}{}_L\,,\olsi{M}{}'_L
\end{array}}
}\hspace{-20pt}%
(4\pi)^2 i^{L-\olsi{L}}
Y_{L}^{M_L}(\Omega_{\uk})^{\ast}
Y_{\olsi{L}}^{\olsi{M}_L}(\Omega_{\uk})
Y_{L'}^{M'_L}(\Omega_{\ur})
Y_{\olsi{L}{}'}^{\olsi{M}{}'_L}(\Omega_{\ur})^{\ast} \times \\
&  h_{S,M_S;L,M_L;\alpha}^{(S',M'_S;L',M'_L;\alpha')}(r,k)
\, h_{S,M_S;\olsi{L},\olsi{M}_L;\alpha}^{(S',M'_S;\olsi{L}{}',\olsi{M}{}'_L;\alpha')}(r,k)^{\ast} = 4\pi \hspace{-10pt} \sum_{%
{\scriptsize\begin{array}{c}
L\,,L'\\
M_L\,,M'_L
\end{array}}
}\hspace{-10pt}%
\left\lvert h_{S,M_S;L,M_L;\alpha}^{(S',M'_S;L',M'_L;\alpha')}(r,k) \right\rvert^2\,.
\end{align}
Therefore, we can write:
\begin{equation}\label{eq:CF_versionB}
    C^{\alpha}(k) =\int 4\pi r^2 dr \, S(r) \, \frac{1}{(2s_1+1)(2s_2+1)} \sum_{S,M_S}\sum_{S',M'_S,\alpha'} \hspace{-10pt} \sum_{%
{\scriptsize\begin{array}{c}
L\,,L'\\
M_L\,,M'_L
\end{array}}
}\hspace{-10pt}%
\left\lvert h_{S,M_S;L,M_L;\alpha}^{(S',M'_S;L',M'_L;\alpha')}(r,k) \right\rvert^2\,.
\end{equation}
Equation\,\eqref{eq:hSMSLMLtoJLS} allows one to simplify part of the sums in Eq.\,\eqref{eq:CF_versionB}
\begin{equation}
\sum_{M_S\,,M'_S,M_L\,,M'_L}
\left\lvert h_{S,M_S;L,M_L;\alpha}^{(S',M'_S;L',M'_L;\alpha')}(r,k) \right\rvert^2 = \sum_{J}(2J+1) \left\lvert h^{J}_{LS\alpha;L'S'\alpha'}(r,k) \right\rvert^2\,,
\end{equation}
\textit{i.e.},
\begin{equation}
    C^{\alpha}(k) = \int 4\pi r^2 dr\, S(r)\, \frac{1}{(2s_1+1)(2s_2+1)} \sum_{L,L',S,S',J,\alpha'} (2J+1)\left\lvert h^{J}_{LS\alpha;L'S'\alpha'}(r,k) \right\rvert^2\,.
\end{equation}
We recall that the functions $h^{J}_{LS\alpha;L'S'\alpha'}(r,k)$ have been defined in Eq.\,\eqref{eq:hJLS}. Now, using together the following identities:
\begin{subequations}%
\begin{align}\label{eq:relations-JLS-s1s2}
\sum_{J}\sum_{S=\lvert s_1 - s_2 \rvert}^{s_1+s_2} (2J+1)\epsilon_{JLS}& =(2s_1+1)(2s_2+1)(2L+1)\,,\\
    \int 4\pi r^2 dr \, S(r) \sum_{L}(2L+1)j_L(kr)^2 & = 1\,,
\end{align}
we can write:
\begin{equation}
    \int 4\pi r^2 dr  \, S(r) \sum_{J,L,S}\frac{2J+1}{(2s_1+1)(2s_2+1)} \, \epsilon_{JLS} \, j_L(kr)^2 = 1\,,
\end{equation}
\end{subequations}
so that:
\begin{equation}\label{eq:General-CF-JLS}
    C^{\alpha}(k) = 1 + \int 4\pi r^2 dr\, S(r)\, \frac{1}{(2s_1+1)(2s_2+1)} \sum_{J} (2J+1) 
    \hspace{-10pt} \sum_{L,L',S,S',\alpha'} \hspace{-2pt} 
    \left\{ \left\lvert h^{J}_{LS\alpha;L'S'\alpha'}(r,k) \right\rvert^2 - \delta_{LL'}\delta_{SS'}\delta_{\alpha\alpha'} \epsilon_{JLS}\epsilon_{JL'S'} j_L(kr)^2 \right\}\,.
\end{equation}
This is the final and general expression for the CF. As can be seen, the multiplicity of each contribution is dictated by $J$, as it should, and weighted by the factor $1/((2s_1+1)(2s_2+1))$. For some cases, it can be more convenient to transform this expression into one where the coefficient $2L+1$ appears more transparently, and this can be achieved by means of Eq.\,\eqref{eq:relations-JLS-s1s2}.

For the particular case we have studied in this work, we have the channels $\alpha = 1$, $2$, and $3$ [$\Xi^{\ast0}K^-$, $\Xi^{\ast-}\olsi{K}{}^0$, and $\Omega^-\eta$], having $L=0$ and $S=3/2$, and $\alpha=4$ and $5$ [$\Xi^0 K^-$ and $\Xi^- \olsi{K}{}^0$], with $L=2$ and $S=1/2$. We have computed the CFs for $\alpha = 1$, $2$, and $3$ exclusively, so that $s_1=\frac{3}{2}$ and $s_2=0$, and we have considered only the $J=3/2$ contribution. Applying this into our general formula Eq.\,\eqref{eq:General-CF-JLS}, we see that the structure of Eq.\,\eqref{eq:correlation} is recovered, except for the necessary generalization to take into account the production weights, as discussed before.

\end{document}